# Intelligent Environmental Empathy (IEE):

## A new power and platform to fostering green obligation for climate peace and justice

**Saleh Afroogh[1]*, Ali Mostafavi[2], Junfeng Jiao[3]**

*1.Urban Information Lab, The University of Texas at Austin, Austin, TX 78712, 4. United States.* Saleh.afroogh@utexas.edu

*2. Associate Professor, Zachry Department of Civil and Environmental Engineering, Texas A&M University, 199 Spence St., College Station, TX 77840, USA; e-mail:* amostafavi@civil.tamu.edu

*3. The School of Architecture, The University of Texas at Austin, Austin, TX 78712, United States.* jjiao@austin.utexas.edu

** Correspondence: Saleh.afroogh@utexas.edu*

### Abstract

In this paper, we propose Intelligent Environmental Empathy (IEE) as a new driver for climate peace and justice, as an emerging issue in the age of big data. We first show that the authoritarian top-down intergovernmental cooperation, through international organizations (e.g., UNEP) for climate justice, could not overcome environmental issues and crevices so far. We elaborate on four grounds of climate injustice (i.e., teleological origin, axiological origin, formation cause, and social epistemic cause), and explain how the lack of empathy and environmental motivation on a global scale causes the failure of all the authoritarian top-down intergovernmental cooperation. Addressing all these issues requires a new button-up approach to climate peace and justice.

Secondly, focusing on the intersection of AI, environmental empathy, and climate justice, we propose a model of Intelligent Environmental Empathy (IEE) for climate peace and justice at the operational level. IEE is empowered by the new power of environmental empathy (as a driver of green obligation for climate justice) and putative decentralized platform of AI (as an operative system against free riders), which Initially, impact citizens and some middle-class decision makers, such as city planners and local administrators, but will eventually affect global decision-makers as well.

## 1. Introduction

For the five decades since the establishment of the United Nations Environment Programme (UNEP) environmental issues have been growing increasingly intricate and abundant on a global scale. Anthropogenic climate change, which is a catalyst, for transformations has been gathering momentum despite whatever actions humans may take in the future. According to UNEP's sixth and latest Global Environment Outlook (GEO-6) report, environmental conditions are deteriorating globally and "the window for action is closing" [1]. As of 1970, the global mean surface temperature has increased by 0.17°C per decade on average, compared to 0.07°C in 1880 [2]. A potential impact could be the doubling in frequency of climate-related events since 1980 [3]. Furthermore, the modern world is not only off-track to achieving green and sustainable environmental goals [1], but also, in practice, the technological origins of the current environmental crisis are being developed rapidly.

UNEP's administrators also suggest that reversing those negative trends and restoring human and planetary health requires immediate action and stronger "international cooperation". It is usually presupposed that "international cooperation" is a necessary and practical platform to address environmental crises. UNEP's history of action shows that, first, by "international" they mean primarily the world's states and governments. Certainly, the goal and leadership of UNEP are not limited to intergovernmental cooperation, and is expanded widely between its non-state partners as well. However, its landmark achievements so far—such as Bonn Convention on Migratory Species, with 116 member states, in 1979; establishment of Intergovernmental Panel on Climate Change (IPCC) in 1988; Basel Ban Amendment, Hazardous waste exporting, which has been ratified by EU and 70 countries in 1995; Launch of Climate and Clean Air Coalition to Reduce Short-Lived Climate Pollutants, which is initiated by 6 states in 2012, and now includes 34 states and EU—clearly show the top-bottom approach in its leadership in addressing climate justice; that is, states in the world and intergovernmental "cooperation" play a primary and central role in their course of action. Secondly, it is historically observable that by "cooperation" they mean a type of democratic collaboration in peaceful and sincere situations between states/governments in the world, without political interruption or egoistic competition.

However, it is statistically shown that such top-down and intergovernmental democratic and peaceful "collaboration" could not overcome the environmental issues and crevices so far, and according to UNEP's report in June 2012, the international community has agreed on 90 environmental goals, but only four have made significant progress. Several reports also show that the goals of the Paris Agreement are not being met by countries [4]. The problem of free riders and misusing climate change as a battle of political and economic competitions casts doubt on the sufficiency or efficacy of intergovernmental and democratic cooperation as a practical solution for environmental issues [5],[6],[7]. Ghimire et al. [8] show that the environmental issues would not be solved in the absence of climate justice, and Goodman [9] examines the relationship between climate justice and the more general problem of global justice, while global justice, in turn, relates to economic and political rivalry between states/governments in the worlds. Such conflicts between the platform of peaceful and non-rival environmental intergovernmental cooperation, and pre-conditionality of climate and global justice (that is, a rival factor) for environmental goals, make it difficult to harbor perfect hope or trust in intergovernmental democratic peace about environmental issues. Since the top-down intergovernmental approach did not come up with favorable outcomes, environmental scholars and activists were led to think of new bottom-up, multi-central or decentralized approaches to environmental issues.

The abovementioned complexities, particularly free riders issues and lack of transparency, cast doubt on the logic and efficacy of the current plans and cooperation, and would negatively affect the environmental motivation in practice between individuals, private sectors, and governmental partners. In addition, data defects and knowledge gaps in climate justice, and green policy making would be important as well. These all call for a new complex system based on big data analysis and a new platform for policy making, implementation, and surveillance [10], [11], [12]. [13] argues for the need for new powers for the UN Environment Programme, as well as some new intelligent platforms to provide a new power and obligation toward environmental guidelines and goals.

In this paper, we propose Intelligent Environmental Empathy as a driver for environmental motivation and green obligations. We explicate the possibility of AI and empathy as a source of new platforms and powers that are needed for climate justice and argue that the intersection of AI and empathy provides us with potential new power for green obligations. The four sections of this paper proceed as follows: section 2 begins with some social political reflections on climate peace in natural status and the problem of climate justice in current civil societies. In section 3, we elaborate on two origins and two causes of climate justice as an emerging issue in recent decades. Section 4 discusses empathy as the significant missing component of the current top-down intergovernmental cooperation, and the need for empathy in the new button-up operational approaches to overcome climate justice issues and some related problems such as the problem of free riders. In section 5, we propose a model of Intelligent Environmental empathy IEE for climate peace and justice at the operational level. This model is empowered by the new power of environmental empathy and decentralized green AI platform; and is going to affect citizens and some middle operational decision makers, such as city planners and local administrators.

## 2. From climate peace in the natural state to climate justice in civil society

Climate justice/injustice is a modern issue in civil societies, while weather conditions and climate factors (e.g., water and air) were not essentially subjects of justice/injustice in the pre-modern era or what is called the “state of nature”. John Locke, in the second of the *Two Treatises of Government*, explains that humans were initially in a state of nature wherever there was lack of government, and people (who were not citizens) acted and reacted based on independent and arbitrary institutions. People in the state of nature still have three basic rights of life, liberty, and property, although in this natural and apolitical situation, there is no government or a common superior to judge between them. Hobbes, in *Leviathan*, depicts the state of nature based on two features—fear of scarce resources and selfishness—that tend to result in conflict. People want and fight over the same natural resources, and life will be poor, solitary, nasty, brutish, and short.

However, the state of nature is unstable, and people are subject to threats of physical harm due to a few ones who seek to live by force and violate natural rights. Moreover, people, in the state of nature, could not pursue goals that required stability and cooperation and were exposed to aggression while they lacked settled laws and impartial judges. So, people decided to become citizens and establish a government by giving up some of the natural rights to a government/state to protect their basic rights. They gave up their power to protect themselves and punish aggressors. In return, the government ensures their safety from any harm safeguards their belongings and fosters a stable environment for them to engage and collaborate with others. It is crucial for governments to be attentive, to the requirements and aspirations of their citizens while safeguarding each individuals rights and freedoms. Hobbes explains rational egoism as a social contract to make peace. That is, fear motivates people to seek peace and stability, and reason leads them to seek peace through a social contract and establishing a Leviathan, which includes a governor/sovereign and nation/citizens.

Hume, in *A Treatise of Human Nature*[14], explains justice as a constructive feature of modern government that comes out of people's fear of scarcity and limitation in natural resources and is led by rational egoism. Moreover, he argues that justice was not in the natural status of human beings. He states that if nature abundantly fulfilled all our needs, wants and desires, there would be no place for the conflicts of interests that justice presupposes, and no use for the distinctions and boundaries relating to property and ownership that "at present are in use among mankind" [14]. For him, scarcity, fear, and selfishness entail the construction of justice, while abundance in resources or benevolence "make justice useless".

Regarding climate peace and justice, Hume argues that human beings were in the climate peace situation in the natural state, where there was no fear of scarcity of natural resources. He argues that property distinctions disappear when enough commodities are available to satisfy all the desires of men. "We can see this with regard to air and water, though they are the most valuable of all external objects; and we can easily conclude that if men were supplied with everything as abundantly as they are with air and water [..] justice and injustice would be unknown among mankind"[14]. Accordingly, he states that justice derives from human selfishness, resource scarcity, and limited generosity among human beings. Hume explains that resource scarcity plays a central role in conflict and justice, and climate factors such as air and water were not subjects of justice in the pre-modern era due to the abundance of resources.

Locke says the same thing in the second of *Two Treatises of Government* [15]. He believes that air and water (or more accurately, the origins of water, like underground sources or rain clouds) were not a property at all in the pre-modern era, because they were not the outcome of industrial or labor productivity, and they were only some common goods given by generous nature. In the pre-modern era, there was an abundance of natural resources and there was no climate conflict, and these climate factors were not subject to justice or injustice. He states that the world and the planet are given to humans in common, which is necessary for survival and enjoyment of life. Seizing too much property leads to waste. Moreover, there should be enough and as good for others to appropriate. If the world and the planet are common property, it is wrong to appropriate more than one's fair share. He also believes that we can cultivate and own property by industrious and rational values so that we add to the natural and common resources, and cultivation increases economic growth. Therefore, Governments might institute rules governing property and distribution. This cultivation and improvement should not be "at the expense of any other man". However, he believed that climate was not a kind of property, at least until the 17$^{th}$ century when he lived. He argued that in the case of water and air, "there was still enough (and as good) left for others" if we use it even more than usual [15]. So, water and air were not considered property for Locke, because there was no scarcity or limitation by using them. He states that "Nobody could think he had been harmed by someone else's taking a long drink of water if there was the whole river of the same water left for him to quench his thirst"[15]. Furthermore, he holds that even land, if there is enough for all to use even inappropriately, cannot be considered property.

## 3. Climate injustice as an emerging issue in the age of big data

As we move into the contemporary era, climate items (e.g., water and clean air) have become more scarce, valuable, and important to humans, and have become increasingly recognized as crucial property in human life, which might give rise to conflicts. Accordingly, the problem of climate justice/injustice has emerged as a modern problem in human society. Today climate and environment are generally subject to justice/injustice, not only as an international problem but also as an intergenerational issue. We are responsible for the next generation as well; and the previous generations in different countries and regions have had different negative/neutral/positive contributions to the status of climate and environmental issues, for which they should be responsible [16], [17]. All these issues make the problem of climate justice/injustice still more complicated. In what follows we elaborate on four major factors (i.e., two origins, one formation causes and one epistemic cause) of the emerging issue of climate justice/injustice. (See figure 1)

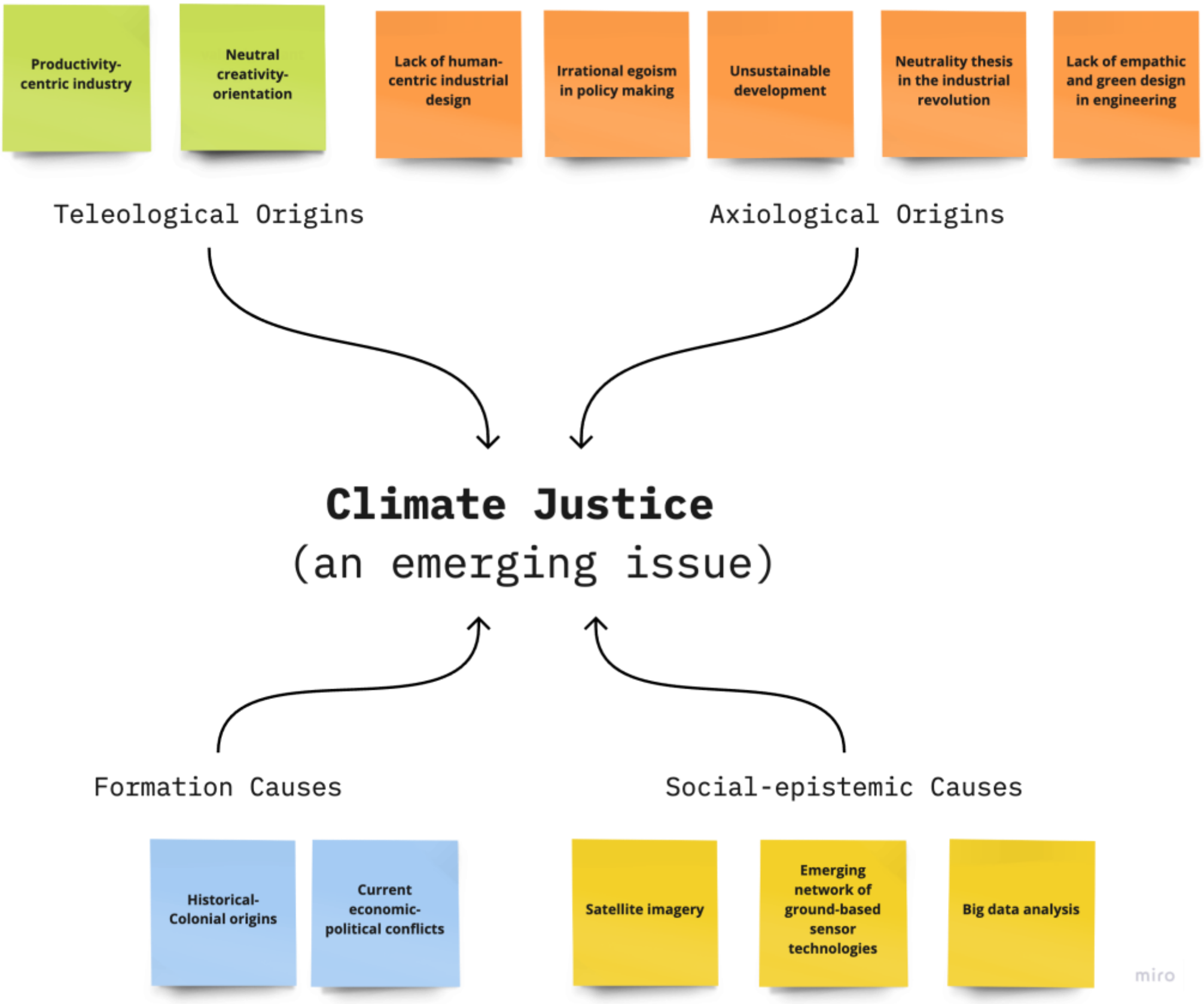

**Figure 1: Climate justice as an emerging issue in the age of big data**

### 3.1. Teleological origins

The first origins of the environmental and climate justice/injustice issues in the contemporary era trace back to a teleological turn during the industrial revolution in the eighteen and nineteen centuries. In spite of the Aristotelian teleological metaphysics that recognizes a type of ultimate moral goal in nature, including human nature and actions, the modern industrial revolution, in eighteenth and nineteenth centuries, turned away from the Aristotelian teleological approach, and began a new path based on human features such as desirability and creativity [18], [19]. According to this new metaphysics, nature (including human nature) is neutral and lacks any teleological features, and human actions are supposed to be human-desire-based; that is, the goals are not already determined (such as certain moral goals) but it is human desires that set the goals. This new approach entails, among other things, two major goals for industrial modern life: neutral creativity-orientation and productivity-centric approach. Thus, regarding the former, whatever is creative and desirable for the stakeholders and the developer would be considered "good", regardless of any external considerations such as moral hazards, environmental considerations, fairness and justice issues, etc. (this is also called the neutrality thesis [20], [21]); and based on the latter the proclivity of the good things would be desirable in the new modern industrial market. Therefore, the modern industry has been recognized, in the past two centuries, as a neutral approach that aims maximum desirability, creativity, and productivity. Such huge neglect of the most basic values in human life has been considered one of the main causes of environmental damages and issues of climate injustice that we are dealing with today.

### 3.2. Axiological origins

The second origin of the current environmental and climate justice/injustice issues is the presupposed axiological issues during the industrial revision in the eighteenth and nineteenth centuries. The neutrality thesis regarding an industrialized modern approach has been undermined in the mid-20th century as a result of the two destructive world wars, and many ethical approaches to engineering emerged over the past few decades, although these changes came too late to prevent problems like environmental injustice and climate change that we face today. Lack of human-centric design in the early modern industrial approach has made the infrastructures of modern life value-ignorant and unsustainable during the centuries[22]. Industrial movement during those centuries was human-desire-based, which is a Hobbesian (and Humean) concept [23] and is diametrically opposed to the human-centric design that mainly pays attention to humanistic values (not desires) and has originated from Kant’s deontological metaethics. Reason and rationality in the modern industrial approach also had been considered just in an instrumental sense [24], [25] that lacks any necessary linkage to moral, environmental, and justice considerations. That is, an agent (such as an industrial developer, designer, or provider) would be rational only if they choose the best method or algorithm to meet their desires, which was maximum creativity and productivity in the market, regardless of external moral or environmental considerations. This approach was instrumentally rational, although today it is observable and intuitive that it was substantially a type of irrational egoism in those generations, which entails unsustainable infrastructure and development that leads to the current dramatic environmental issues and climate injustice. Such a neutral and value-ignorant, unsustainable, and irrational egoistic approach has been considered another major cause of environmental issues, particularly climate injustice/justice [26].

### 3.3. Formation causes

The formation cause of the climate justice/injustice issue is two-fold: first, historical colonial origins, and second, the current economical-political conflicts. According to [27] “Climate politics is without doubt a postcolonial issue and the concept of climate justice emerged out of a postcolonial critique of global climate policy”; in fact, climate justice, in one sense, means addressing the colonial extractionist practices that have led to climate change in the first place[28]. Any movement toward climate justice should be a type of de-colonizing approach [29]. For a long time, colonizing countries exploited natural resources of colonial countries and established their industrial infrastructures. Moreover, cheap human resources in poor countries have motivated big companies to establish their main manufacturing branches in those countries, which entails increasing $CO_2$ in those regions. Porter et al. [28] explain that in settler-colonial Australia, all climate change research and action took place on Indigenous lands unceded by the Australian government. Falzon and Batur [30] examine the influence of racist colonialism to environmental damage and climate injustice. A detailed explanation was given of colonial histories of the Pacific islands, including Nauru and Banaba (now part of Kiribati and two French colonies in the past) as places of extensive environmental damage to the benefit of the colonizers, and for the detriment of the colonized people living there. Pacific Islanders are actively being neglected by the world, and due to human industry and energy production, $CO_2$ and other greenhouse gases are increasing, causing the islands of the Pacific to disappear.

Rice et al. [31] investigate how these historical colonizing structures of oppression promote mainstream responses to global ecological catastrophe to systematically promote spatial, socioeconomic, and economic segregation. Moreover, it is shown that "Climate politics today is also the colonization of the future" and make the policy makers responsible for the next generations [27]. In fact, along with the fact that the scarcity of climate factors like clean air and water increased and made them worthy and crucial property for human life, increased the economic and political conflicts over the ownership of these natural resources, and directly caused climate change and injustice issues in the contemporary era and the future.

### 3.4. Social epistemic causes

The last causal factor of the emerging issue of climate justice/injustice consists in social episteme causes. This is not a formation cause. That is, it does not naturally or physically cause climate change and injustice issues. It rather enables human beings, epistemically, to figure out much more clearly the issue of climate justice, and distinguish it as a newly emerging problem of human modern life. Big data and technological revolutions have greatly impacted climate science as a data-intensive field. Big climate data analytics implementations have been focused on climate change as an emerging topic and comprehensive research has been conducted on a number of topics. All of them informed today's humans to acknowledge and pay attention to the emerging problems of climate justice/injustice [32].

The big data collections through satellite imagery and the internal circulation of information through social networks such as twitter have played a central role in informing and warning humans of climate issues worldwide [33], [32]. Earth observation technology has greatly benefited global climate change research by providing biological, physical, and chemical parameters on a global scale. Researching climate change can take advantage of the four V's (volume, variety, veracity, and velocity) of Earth observation data. Guo et al. [34] study the advances in climate change studies using Earth observational big data, combined with terrestrial observational sensitive variables (such as lakes, glaciers, radiation, and vegetation). The use of big data from earth observation has enabled us to identify and confront global climate change challenges. Faghmous and Kumar [35] developed a theory-guided approach to data science and interpreted big climate data in order to gain an accurate insight into climate change. Moreover, Data for Climate Action campaign is a type of "data for good" institution, which consists of a set of collaborative efforts that use user data to address climate change. These companies supply their proprietary big datasets to NGOs, international agencies, and development agencies in order to assist them in recognizing and addressing climate and environmental problems [11]. However, there has not been much impact of data science (compressed by its effect on advertising and e-commerce) on addressing climate justice/injustice issues, and this is closely related to the problem of data injustice/justice, to which we come back in the next section [35].

## 4. Empathy as a missing component in top-down intergovernmental cooperation for climate justice

In the five decades since The United Nations Environment Programme (UNEP) was established UNEP has made strides in advancing environmental and climate science as well, as legislation. One notable achievement came in 1988 when UNEP collaborated with the World Meteorological Organization to establish the Intergovernmental Panel, on Climate Change (IPCC). Additionally UNEP has been advocating for policy reforms that discourage funding of industries causing pollution. As part of its efforts, it has also advocated working with China to green its rapid industrial initiatives, such as the Belt and Road Initiative. However, there was not a "globe's green watchdog" that forced governments to follow green rules. A major concern raised by developing countries during the Stockholm Conference in 1972 was the impact of such an organization on industrialization in their countries [13].

In spite of the United Nations' flagship Sustainable Development Goals' rising priority, progress is almost nonexistent. There is not a system for ensuring compliance with Sustainable Development Goals. According to the Initial NDC Synthesis Report published in 2021 on countries progress towards commitments under the 2015 Paris climate agreement, just nearly 30% of global greenhouse gas emissions have been reported from 75 Parties with new or updated NDCs, while the agreement includes almost 200 countries. Moreover, the UK and the European Union were the only two out of 18 largest emitters to present a new NDC in 2020 that contained a significant increase in its target for reducing GHG emissions [36].

As the most important top-down cooperation, international cooperation was not successful in achieving the UNEP green goals, and it is suggested that we need new power and platform to enable us to achieve the goals [13]. In the next sections, we will explain that the lack of empathy entails the lack of environmental motivation and cause the problem of free riders in the intergovernmental level; and we will argue for empathy as the new complementary power to increase the environmental motivation to the extent of green obligations and propose AI (i.e. empathic AI or AI as empathy generator) as the new platform to address the problem of free riders as a major barrier to achieving the goals.

### 4.1. Empathy as a new source of environmental motivation and green obligation

To achieve a green obligation upon governmental parties, it is suggested that we need a new radical source of power to create a green motivation to support intergovernmental cooperation for environmental goals. According to [13] UNEP member nations are considering enhancing the organizations authority to function as a body that ensures compliance. It is likened to the World Trade Organization in terms of its ability to impose penalties on countries, for non compliance, with its agreements. However this proposal has been deemed as extreme. While such a perspective may have been justifiable two decades ago it is now imperative to take actions, including empowering UNEP to fulfill its responsibilities given the severity of climate change we are currently facing.

However, it does not take account of the fact that every radical punishment or legislation might entail radical resistance or increasing free riders between the parties. It might end up in radical conflicts, which could then result in a *reductio ad absurdum*. Needless to explain that any radical conflicts such as war and military involvement can destroy the environmental achievements that are supposed to be protected.

The hard radical punishment and legislation not only is impractical but also might imply certain negative effects for environmental goals. We believe that the obligation can be fulfilled in the absence of radical legislation or punishment. From an action-theoretical and psychological viewpoint, every human action is processed by a complete will and motivation that end up in decision and action. So, forcing people or stakeholders through some external forces is not the only path toward fulfilment of green obligation. Green obligation can be fulfilled by an environmental motivation that originates form cognitive attitudes (such as rational choice theory) or affected attitudes such as empathy.

Empathy as an affective source has been considered for moral and legal obligation [37], [38], and accordingly, we can consider it as a new power and source for the green obligation as well [39], [40]. Empathy refers to the capacity to comprehend and sympathize with the emotions of others. When empathy is absent it can result in a disregard, for the well being of others encompassing both living beings and the environment. Individuals who lack empathy may exhibit an inclination to participate in environmental activities or address issues related to climate justice. The lack of empathy can lead to a lack of environmental motivation and ultimately result in a failure to fulfill green obligations for climate justice. Individuals who lack empathy may not feel personal and social responsibility towards environmental issues and climate justice, which can make it difficult for them to understand the need to take measures to safeguard it. They may not experience an existential sense of duty towards preserving the natural world, which is essential for environmental motivation to made green decisions for climate justice.

Moreover people (including government policymakers) who struggle with empathy might not fully grasp the consequences of their actions, on the environment. This can result in a lack of motivation to modify their behavior since they may fail to recognize the impact they have on the environment. It is vital to acknowledge that individuals who lack empathy towards nature may not perceive its value or significance. Therefore it is crucial to cultivate an understanding of the importance of empathy in addressing climate justice and fostering a future.

The inefficiency of collaboration can be attributed to a core deficiency, in empathy. Governments in a Westphalian order frequently prioritize immediate economic growth over environmental considerations, as they are more interested in short-term benefits than long-term sustainability. This deficiency of empathy towards the natural world and future generations leads to a disregard for the significance of tackling environmental issues, ultimately causing governments to fail to work together. Without empathy and a willingness to address the concerns of the environment and future generations, intergovernmental cooperation cannot succeed in mitigating climate injustice.

The biggest greenhouse gas emitters in the world, who happen to be the most economically advanced nations, prioritize economic growth over environmental protection. Despite being aware of the negative impacts of their actions, the competition and economic interests among them discourage significant efforts to change this trend. As a result, these countries remain among the largest carbon emitters globally. An study published in Carbon Brief in 2021, analyzes the historical responsibility of countries for CO2 emissions since 1850. The quantity of emitted CO2 since the industrial revolution is strongly linked to the present 1.2C temperature increase. In the time period from 1850, humans have injected roughly 2,500 billion tons of CO2 into the atmosphere, leaving a meager 500GtCO2 in the remaining carbon budget to remain below a warming of 1.5C. Historical emissions mostly come from the United States making up 20% of the global emissions. China follows with 11%, Russia with 7%, Brazil with 5% and Indonesia with 4%. Notably larger European nations with histories like Germany and the United Kingdom contribute 4% and 3% respectively to the global emissions. This analysis considers consumption based emissions accounting to reflect trade in goods and services that're carbon intensive. The relationship between emissions and temperature rise is measured through a called "transient climate response to cumulative emissions" (TCRE) which estimates a warming effect of around 1.65 degrees Celsius per every trillion metric tons of carbon (or equivalently about 0.45 degrees Celsius, per every trillion metric tons or gigatons CO2). [41]

To address these issues, a new bottom-up approach to climate peace and justice is needed. This perspective highlights the significance of responsibility and community driven movements, in promoting sustainability and fairness. By cultivating a concern for nature and all living beings people can find inspiration to actively safeguard the world. This may involve making lifestyle changes supporting environmental initiatives and advocating for government policies that prioritize sustainability.

A grassroots approach to climate justice gives priority to community efforts, in addressing challenges rather than solely relying on government intervention. This approach acknowledges that meaningful change often originates at the level, where individuals and community organizations proactively tackle issues. An example of this approach is the grassroots movement against single-use plastics. People have taken small-scale actions such as reducing their plastic use, advocating for local plastic bans, and pressuring companies to reduce their plastic packaging. These actions have gained momentum, leading to larger changes, such as nationwide bans on certain single-use plastics.

By fostering empathy towards the environment and other living beings, individuals can be motivated to take action to protect the natural world. For example, someone might choose to reduce their meat consumption after learning about the environmental impact of animal agriculture or switch to public transportation after understanding the harm caused by individual car use. In summary, the bottom-up approach recognizes the power of individual and community action in promoting environmental sustainability and justice, as working together towards a common goal can have a significant impact on creating a more sustainable and just future.

### 4.2 Westphalian world order, green Leviathan, and the problem of free riders

Moreover, any radical legislation or political coercion probably ignores the feature of rivalry between countries in the Westphalian order. The Westphalian order, as opposed to the indigenous order, refers to the modern state-centric world order, in which states are the smallest political components of the world order, and the state's supreme political authority would be part of the established legal framework in international law. These states are independent and responsible to protect the rights and interests of their own nations and citizens. The current state-nation system originated in Hobbesian Leviathan theory [42], [43]. Leviathan [44] is a metaphor that represents Hobbes's ideal government. Leviathan is necessary to preserve peace and prevent civil war. It is formed by nation and sovereign, who is the head of the leviathan. Peace can be achieved only by building a leviathan through a social contract. The major role of a leviathan is to consider citizens' rights and protect and increase their interests in conflict and rivalry with other leviathans in the world. Accordingly, on the one hand, (1) rivalry in economy and politics is an essential feature of the current nation-state political system in the world. On the other hand, (2.1) any effective solution for climate injustice is necessarily related to solutions for global injustice [45] [9]; (2.2) global justice requires Leviathan to stop political and economic rivalry irrespective of the problem of justice and requires any specific Leviathan to pay back all the exploited money which is gained during the colonial history to the colonized countries; (2.3.) 2.2 is contradicted by 1, which states that rivalry is an essential feature of the current nation-state systems, and is contradicted by 2.2., which requires putting an end to political and economic rivalry irrespective of the problem of justice.

This contradiction might be one of the origins of the failure of any intergovernmental cooperation and the problem of free riders in environmental and climate issues as well. That is, governments in the Westphalian system, as independent leviathans, cannot prioritize other nations' environmental interests or even common global environmental interests to direct economic interests of their nations. In other words, any government, by definition and nature, would prioritize direct economic interests of its nation to empower itself against other leviathans and there might use environmental factors in this rivalry as well, as we see in the case of international water conflicts [46].

Accordingly, the current sate-nation order that forms different leviathans would be practical and effective only if the major desire of leviathan (which comes from the citizens' desires) were environmental considerations as a rivalry factor that could make it more powerful against other leviathans. For example, if all the nations strongly desire environmental goals, it forms a strong desire in a government to follow them, and it makes the leviathan still stronger than other leviathans. That is, making a green leviathan would be possible only if we increase the environmental motivation and create a green obligation in the nations and stakeholders and human factors of any industry and government. Such human modification automatically entails the green leviathan, at least in democratic societies or rational monarchies.

Ignoring the contradiction between Westphalian systems and the requirements of the claimant and global justice also entails free rider problems in international cooperation. To address this problem, we can use green AI as a powerful surveillance platform. Empathy can play a pivotal role both as a new aspect of green obligations and as a feature of green artificial intelligence. Empathy can play an important role in creating a sense of obligation towards environmental goals and promoting collective action to address the free rider problem. Empathy plays a crucial role in fostering a deep sense of interconnectedness between individuals and societies, enabling them to recognize the significance of climate change for both present and future generations. It also aids in placing environmental concerns above immediate economic interests. By actively promoting empathy, we can effectively catalyze a cultural transformation towards a more sustainable future, ultimately fostering greater willingness among individuals and societies to collaborate and actively contribute towards environmental goals. Additionally, the use of green AI can also help address the free rider problem by monitoring and regulating environmental policies and behaviors, and promoting greater transparency and accountability in environmental governance.

### 4.3. Fostering empathy for climate justice at middle operational levels (citizens and city decision makers) vs. in the global level (intergovernmental policy makers)

The approach to building environmental empathy can be either top-down, focusing on the government and global intergovernmental institutions, or bottom-up, focusing on citizens and middle operational levels. The top-down approach to building environmental empathy focuses on the government and global intergovernmental institutions as the primary drivers of environmental action. This approach would be challenging for serval reasons.

Firstly, many governments primarily prioritize their own short-term economic and political interests, making it difficult to give immediate attention to environmental concerns that may yield long-term benefits but require immediate sacrifices. In these systems the main emphasis remains on the authority of states and governments are hesitant to relinquish control to bodies.

Secondly, as global intergovernmental organizations, like the United Nations require agreement, from all member nations they frequently need to tackle the underlying factors contributing to issues. This often results in diluted agreements that fail to effectively tackle environmental issues. This challenge further intensifies in matters with significant economic or political implications, such as climate change, where countries may be reluctant to commit to binding agreements that could affect their economic competitiveness or national security.

Thirdly, the absence of robust enforcement mechanisms for international agreements poses a major drawback to the global and intergovernmental approach in addressing environmental issues. Despite the agreements reached by multiple countries, some nations do not fulfill their obligations, which hinders progress in resolving environmental problems. Additionally, the decision-making process in a global approach can be slow and convoluted, resulting in delays in taking action on pressing environmental concerns. This can be particularly concerning in cases where environmental degradation is happening rapidly and demands prompt attention.

In addition, to the concerns mentioned earlier taking an approach also poses the risk of reinforcing power imbalances. More influential nations and corporations could have a say, in decision making processes, which can lead to injustices and inequalities. This can particularly impact less developed nations well as marginalized communities. Additionally, there are instances where the global approach fails to effectively engage the public, impeding progress in addressing environmental issues. This becomes particularly problematic in cases where public involvement is vital for success. Hence, while the global approach can be valuable in tackling environmental issues, it is crucial to address several potential drawbacks.

Another challenge emerges from the absence of a shared understanding or agreement on the definition and scope of environmental issues. This lack of consensus often leads different countries to prioritize varied aspects of the problem. Some may concentrate on reducing greenhouse gas emissions, while others may prioritize biodiversity conservation or water management. In essence, while global intergovernmental cooperation remains vital for tackling environmental challenges, instilling environmental empathy at this level becomes challenging due to the complex nature of these issues and the divergent interests of different governments.

Therefore it is crucial to prioritize the cultivation of empathy, towards the environment at both national scales. This is because when individuals and communities directly witness the effects of degradation they develop a sense of personal connection and duty towards safeguarding the environment. Encouraging people to take action in preserving the environment promoting practices and minimizing emissions on a level can set an example, for other communities eventually leading to a positive global influence. Building environmental empathy at the citizen and middle operation levels can be a more effective way to promote environmental sustainability and justice. (See, Table 1)

However, building environmental empathy at the citizen and middle operation levels can be more effective than at the governmental level. citizens' demands can influence government actions. Should citizens awaken to their environmental responsibilities and vociferously clamor for action, governments may find themselves compelled to embark upon measures to tackle environmental quandaries. Moreover, local authorities possess the power to wield a profound influence over the environment, armed with an intimate comprehension of the idiosyncratic circumstances and predicaments that beset their locales. They can unleash a torrent of policies and programs, designed to curtail waste, champion public transportation, and foster energy-efficient architectural marvels, among myriad other endeavors. By immersing themselves in the tapestry of local communities and enlisting their active participation in decision-making processes pertaining to environmental conundrums, individuals can foster an unwavering appreciation for sustainability and the concomitant repercussions of their actions upon the environment.

**Table 1: Fostering empathy for climate justice: drivers at the operational level and barriers at the global level**

| Fostering empathy for climate justice | Derivers at middle operational levels (citizens and city decision makers) | Barriers at the global level (intergovernmental policy makers) |
|---|---|---|
| 1 | Creating a deeper connection to the local environment and a greater appreciation for protecting it | Governments, typically, prioritize their own economic and political interests, leading to competing interests among different governments that often conflict with efforts to promote sustainability. |
| 2 | Encourages involvement in decision-making processes and increases public engagement. | Global intergovernmental institutions are often constrained by the need to achieve consensus among all member states |

| 3 | Engaging individuals in community-based initiatives | Lack of enforcement mechanisms for international agreements |
|---|---|---|
| 4 | Direct and immediate action can be taken | Decision-making process in a global approach can be slow and convoluted |
| 5 | Promotes a shift towards more sustainable lifestyles and behaviors; and so promotes social justice and equity | More influential nations and corporations have more significant influence in decision-making |
| 6 | Better understanding of local situations and challenges make stronger motivating in citizens | Failure to engage the public effectively |
| 7 | Developing a sense of shared purpose and a greater understanding of the interconnectedness of all living beings | Lack of shared understanding or agreement on the definition and scope of environmental issues, or different countries prioritizing different aspects of the problem |
| 8 | Cities and regions can serve as models for other communities; and fosters a sense of connection to the larger ecological system | Reluctance to commit to binding agreements that could affect economic competitiveness or national security |
| 9 | Builds trust between community members | Lack of trust between nations |
| 10 | Building a sense of ownership, community, and responsibility for the environment | Limited public participation in decision-making processes |
| 11 | Empowers citizens to demand action from government | Limited political will in some nations |

Additionally, the bottom-up approach, which involves citizens and middle-level operations, has several ways of addressing the challenges of building environmental empathy and promoting sustainability. The approach enables direct and immediate action to be taken towards environmental issues by engaging individuals in community-based initiatives such as neighborhood gardens, recycling programs, and local conservation efforts. This involvement creates a deeper connection to the local environment and a greater appreciation for protecting it.

Additionally, the ascend-from-the-bottom tactic fosters the construction of a collective conscience and a joint obligation towards the environment. Through collaborative efforts to accomplish a shared objective, individuals have the potential to cultivate a sense of united determination and a profound comprehension of the interdependence among all sentient creatures. This comprehension stimulates a transition towards more enduring ways of life and conduct, as individuals perceive their own existence as an integral component of a grander ecological network. Moreover, by fostering environmental instruction and consciousness at the grassroots level, people are equipped to champion transformation at elevated tiers of governance. It can include lobbying for more sustainable policies and practices or engaging in political activism to push for greater environmental protections.

Aside from these benefits, the bottom-up approach has additional advantages over the global approach in promoting environmental sustainability and justice. The approach can better address local environmental issues and concerns because local communities often have unique environmental challenges that global initiatives do not adequately address. The approach additionally fosters a sentiment of possession and accountability for environmental matters amid individuals and communities, sparking a cultural metamorphosis towards sustainability that outlasts and permeates more extensively than authoritarian tactics. Ultimately, the grassroots approach cultivates sturdier and more adaptable communities that are aptly prepared to confront environmental obstacles by nurturing collaboration and cooperation at the grassroots level.

One example of the bottom-up approach is Barcelona's Superblocks program which is focused on reducing car traffic and promoting sustainable transportation in specific neighborhoods. Empathy in superblocks works as a powerful tool for promoting sustainable behavior and creating a more connected and livable city. In this initiative, it wields a momentous role, as inhabitants are urged to contemplate the reverberation of their actions on the environment and their personal well-being. The Superblocks program endeavors to kindle empathy by nurturing a sense of collective obligation for forging a more enduring and inhabitable metropolis. This can engender a more nurturing community for the program, as well as enduring transformations in conduct towards sustainability. Furthermore, the program places a premium on foot-friendly and cycle-friendly thoroughfares over vehicular congestion, engendering a more comprehensive and egalitarian urban milieu for all constituents of the community.

## 5. Toward an intelligent environmental empathy at operational level: Intersection of AI, empathy, and climate justice

The power of empathy as a new source to increase environmental motivation would make a path to fulfilment of green obligations, which is needed for climate justice (See figure 2). Climate justice also includes several dimensions, including intergenerational justice [47], international justice [48], and intercontinental (or interregional) climate justice [49], all of which could be considered subgoals and could be addressed by the new power of empathy in AI platform.

**Figure 2: Intersection of AI, empathy, and climate justice**

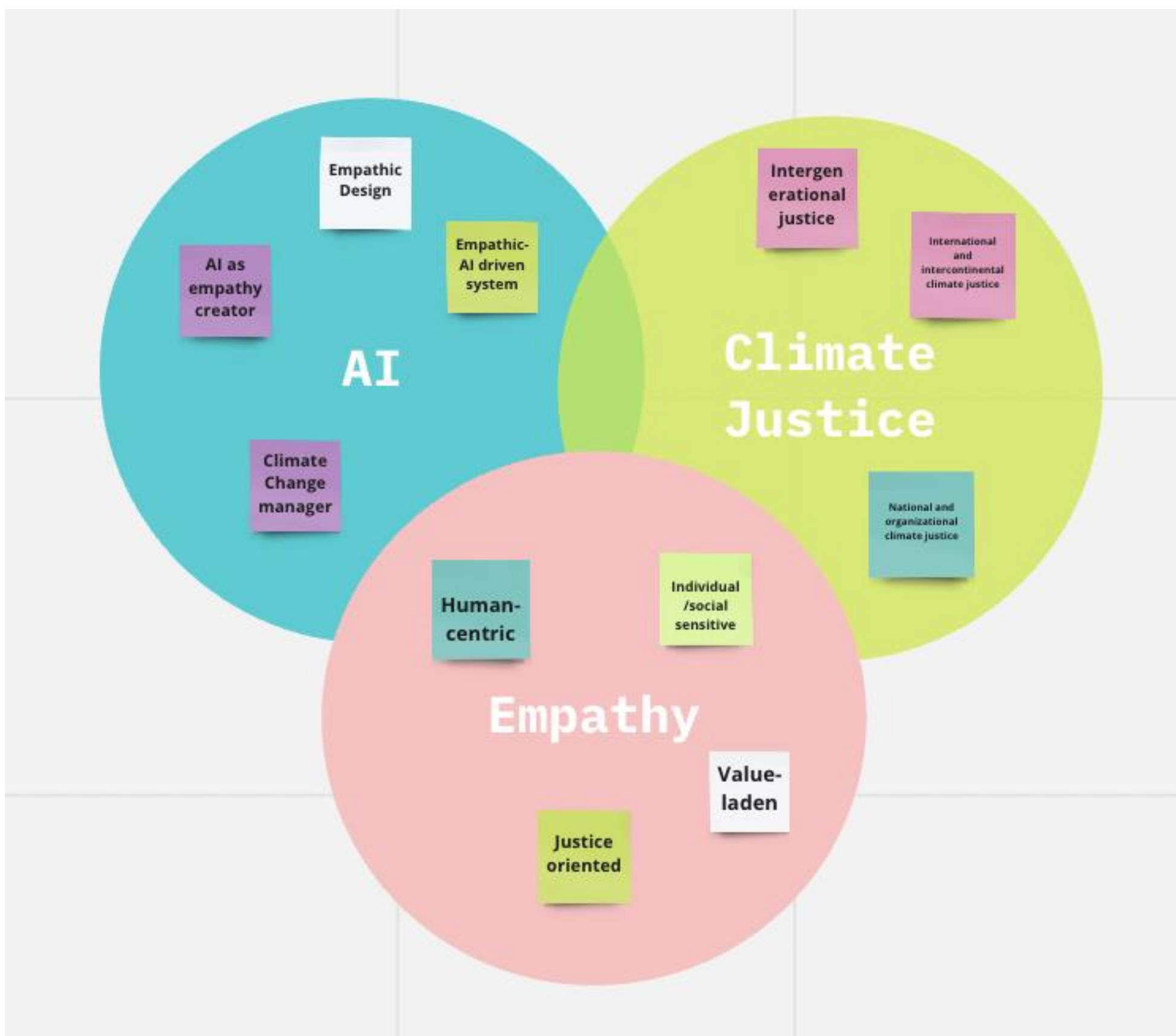

Green obligation, as a commitment to protect the environment and consider the implications of climate justice, requires a strong environmental motivations based on humanistic attitudes. Empathy, in its cognitive, affective, and conative varieties, could function as a rich source of environmental empathy, which in turn would entail environmental motivation. Empathy also plays a pivotal role in human-centric design [20] and justice-oriented design [50]. It also functions as a tool to create value sensitivity both at individual and social levels [51]. In addition, to address the problem of free riders, AI as a new platform could play a central role in green surveillance. AI-driven systems would also empower climate change management in several ways, including automated warnings [52], climate change modeling and forecasting [53], data assimilation [54], digital twin modeling and replication [55], etc.

Furthermore, AI systems possess the capability to bolster a grassroots strategy towards climate justice and foster empathy amidst stakeholders in myriad manners (refer to Table 2). Primarily, data-powered decision-making can empower policymakers to discern patterns and trends linked to climate change, environmental ramifications, and social equity, engendering insightful policy formulation and judicious resource distribution. Secondly, AI systems can expedite community involvement, granting a platform for the voices and necessities of the most impacted communities to be acknowledged and attended to. Thirdly, communication and education campaigns can be developed using AI systems to raise awareness about climate change and promote empathy and understanding among stakeholders. Fourthly, AI systems can employ prophetic modeling to grasp the conceivable ramifications of actions on climate change. Ultimately, AI systems can expedite interactive troubleshooting, fostering empathy and comprehension among stakeholders while propelling superior solutions. To attain a grassroots approach to climate justice, prioritizing collaboration, community involvement, and data-fueled decision-making is imperative. By placing emphasis on empathy and understanding, AI systems can support the development of more effective and equitable solutions to the challenges of climate change.

**Table 2: Some AI systems with the potential to enhance stakeholder empathy for climate justice**

| AI systems | **Primary functions** | **The potentials: How it fosters environmental empathy** |
|---|---|---|
| Climate Visuals | A research-based database of images and videos that can be used to visually communicate climate change impacts to promote empathy and action. | Climate Visuals provides a database of images that promote effective climate communication, helping to foster empathy for the environment and encourage more meaningful climate action. |
| JouleBug | A mobile app that encourages sustainable habits and behaviors through gamification | JouleBug makes sustainable living fun and engaging through its gamified approach, helping users see the impact of their actions on the environment and promoting more empathetic behavior towards it. |
| Urban Footprint | A tool that helps cities make data-driven decisions about sustainability | As an AI system that provides cities with data and insights to make more informed decisions about sustainability. By analyzing data, users can minimize their carbon footprint and better understand the impact of their decisions on the environment and the people. This helps build empathy among city |
| | | leaders and citizens alike towards the environment and its inhabitants. |
| Giki Zero | A UK-based Social Enterprise and B Corp with a primary objective to assist individuals in living sustainably. The team at Giki comprises like-minded individuals who are committed to reducing carbon emissions and helping individuals worldwide to do the same. | Giki offers details on the ecological effects of various items and urges users to select more eco-friendly options, thereby cultivating environmental empathy by means of consumer behaviour. |
| OLIO | OLIO is a platform that enables neighbors and local businesses to share surplus food, rather than throwing it away. It connects individuals with each other and with local stores to share food that is nearing its sell-by date or surplus produce from home gardens. By reducing food waste, OLIO promotes a more sustainable and environmentally friendly lifestyle. | Olio connects users with surplus food that would otherwise go to waste. By making it easier for people to share food and reduce waste, Olio fosters empathy towards those who are less fortunate and helps to reduce the environmental impact of food waste. |

AI systems can increase environmental empathy in citizens by providing them with accessible, real-time data on environmental issues. By employing unconventional, mind-boggling verbs, people can unravel the profound repercussions of their daily choices on the environment and be enticed to embrace more eco-conscious decisions. Furthermore, AI systems possess the remarkable ability to fabricate immersive encounters, empowering individuals to apprehend the far-reaching consequences of environmental issues on communities and ecosystems. For instance, AI can conjure up virtual reality simulations that plunge users into the heart of climate change impacts, leaving them bewildered and enlightened. AI can also personalize environmental messaging based on user data, providing tailored recommendations to help reduce environmental footprints. Overall, AI can enhance environmental empathy by providing data, immersive experiences, and personalized messaging to inspire action and behavior change.

Moreover, AI systems can help democratize decision-making related to environmental issues by providing citizens with access to data and information. Such knowledge has the potential to embolden individuals to discern the paramountcy of their surroundings and apprehend the reverberating repercussions of their actions. By perceiving the environment firsthand and embracing their role as custodians of nature, individuals might cultivate a profound emotional rapport with their surroundings and experience an enhanced obligation to safeguard it. Such active participation has the potential to embolden individuals and foster a profound sense of ownership and accountability towards the environment, thereby engendering a populace that is more involved and well-informed. Additionally, providing citizens with access to data and information through AI systems can help build empathy towards environmental issues and foster a sense of environmental empathy, which is crucial in addressing climate justice issues.

AI systems can aid city planners and local administrators in developing environmental empathy by offering them data-driven insights and decision support tools. By analyzing satellite imagery and other data sources, AI systems can identify the areas of a city most vulnerable to environmental risks and assist with urban planning decisions such as building new infrastructure or redesigning existing buildings and neighborhoods to be more environmentally friendly. AI can also model different scenarios and predict the environmental impact of different development options, allowing planners to minimize environmental harm. Furthermore, AI possesses the ability to scrutinize environmental hazards in the present moment by scrutinizing data from sensors that gauge air quality or water pollution levels. This data can then be employed to avert or alleviate harm to the environment. In essence, AI equips city planners and local administrators with the necessary instruments to make well-informed decisions that take into account the plausible environmental repercussions of their endeavors.

there are similarities and differences between the AI systems designed to build environmental empathy in citizens and those designed for city planners and administrators. Both types of systems aim to build empathy towards the environment, use data analysis and visualization, and provide personalized recommendations. However, the citizen-focused systems aim to encourage individual behavior change through mobile apps and social media, while the city planner and administrator systems focus on systemic change through specialized software and collaboration with multiple stakeholders. Additionally, the citizen-focused systems may use gamification, while the city planner and administrator systems prioritize data-driven decision making and policy recommendations.

### 5.1. A model of Intelligent Environmental Empathy for Climate Peace and Justice

In this section, we elaborate on the conceptual model of IEE for climate peace and justice (See Figure 3). This model is empowered by the new internal power of empathy, and the putative decentralized platform of AI, which enables it to be inclusive enough in dealing with the global problem of climate justice and the free riders problem.

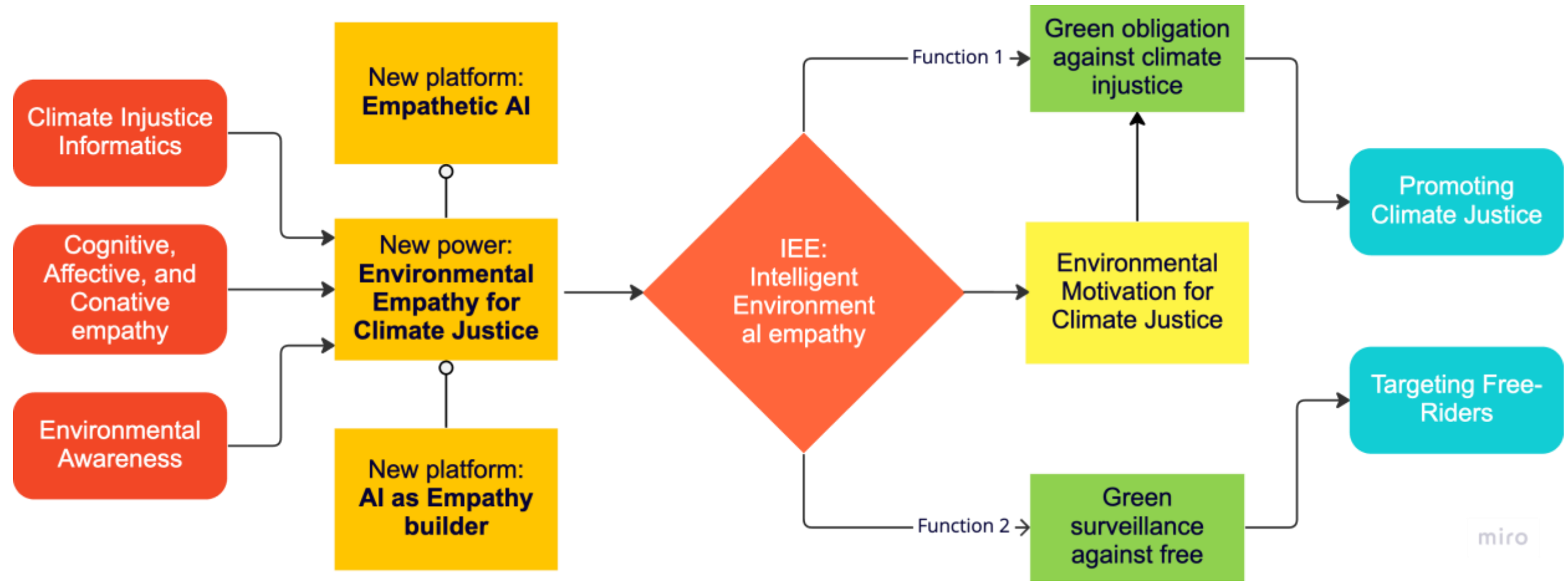

**Figure 3: A Model of Intelligent Environmental Empathy (IEE) for Climate Peace and Justice**

### 5.1.1 Environmental empathy as a new power for climate justice

Climate justice and peace require a new power of environmental empathy. On one side, climate justice and peace have not been achieved appropriately through top-down approaches like intergovernmental cooperation; and on the other side, such an obligation does not sound rational to be archived by some external political or military force (e.g., United Nations Security Council) that might result in radical conflicts, and so increase environmental damage and injustice in the world.

In this model, environmental empathy would function as a strong internal motivation maker, and a driver for environmental motivation, which if provided enough, would entail the green obligation needed for climate peace and justice. Environmental empathy is applicable at all three levels of cognitive, affective, and conative environmental empathy [20]. All these levels are required to be developed and built in three major classes of the private sector, governmental policymakers, and public users. Thus, it could effectively bring up the green obligation for climate peace and justice.

Two major sources to promote and build environmental empathy are as follows: (1) climate justice informatics, which fills the information gap regarding the source of climate change factors, assessment, and evaluation of the responsibilities of different parties regarding their actions, economic effects, charges, and damages of their actions, etc. (2) Environmental awareness, which includes a broader viewpoint than the former, and encompasses qualitative dimensions of climate justice informatics, such as sociological, psychological, and cultural dimensions related to data and their effects on human agents as public users, policymakers, and private sector managers.

### 5.1.2. AI as a new platform: for environmental empathy and against free riders

In this model, AI systems function as an intelligent, self-controlled, and global platform to create and build empathy in stakeholders, based on climate justice informatics, and environmental awareness. For instance, one of the major problems that entail climate injustice and conflicts is lack of enough information about sources of $CO_2$ emissions and other instances of climate change caused by different countries and sectors. This obscurity entails more complicated conflicts, especially in distinguishing the responsibility of each party, sector, agent, and government, for their non-green actions. According to Initial NDC Synthesis Report [36], only 75 countries out of 200 countries have released their environmental data. AI systems, through data assimilation and machine learning methods, can provide a reliable or best possible estimation of data, and it would be beneficial for assessing and evaluating the proposed or claimed data sets. Big data mining, cleaning and verification methods [56] would enable climate change managers to have a better understanding of free riders and climate disaster factors in different regions.

Platforms based on artificial intelligence could also facilitate decentralized approaches to climate change management. The centralized, top-down, and authoritarian policymaking regarding environmental issues has been criticized and considered an unsustainable approach [57],[58]. "Most observers agree that more than 20 years of UN climate negotiations have been a failure. Some argue that the top-down approach is one important reason for this and that a bottom-up approach or more exclusive club approaches would have rendered better results" [59]. Goldthau [60] argues that local and decentralized solutions and newly redesigned infrastructures are necessary to address the energy access and low carbon challenges. Tormos-Aponte et al. [61] examine the connection between climate justice and climate change governance through a case study, demonstrating that climate justice groups are becoming increasingly polycentric as they seek to address environmental problems on a multiscale and the limitations of existing institutional arrangements. In their analysis, they conclude that these polycentric arrangements facilitate the simultaneous influence of environmental movements like those for climate justice across multiple levels of environmental governance, from the local to the global, contributing to an increase in formal institutional polycentricity. Grimaud et al. [62] propose an endogenous model using decentralized equilibrium analysis for climate change mitigation. To address climate change issues, different levels of government must clarify their responsibilities, and decentralizing responsibility would require decentralized operation and management, and Industry 4.0 smart that is employed by AI plays a central role in the decentralized approach [63]. Henceforth, the European Union, driven by its unwavering resolve to curb global climate change, instigated the grandest and most audacious venture of a emissions trading program ever witnessed in 2005. The EU Emissions Trading Scheme, boasting unique features, sets itself apart from the conventional paradigm. This system is relatively decentralized, with each member state being responsible for setting targets, and has some control over permits, verification, and enforcement [64].

AI correlation with empathy is twofold. It can function as empathy builder in stakeholders as well as a central constituent of empathetic AI-driven system for climate change management (see figure 4). Using global connection through intelligent and global AI systems would increase environmental awareness, and a decentralized approach would allow each stakeholder to take responsibility in a more decentralized manner. The environmental awareness and understanding of decentralized responsibility are the first step of creating and improving environmental empathy in human agents. Platforms based on artificial intelligence facilitate the design of multitask systems that can display, through simulation methods, each public user, governmental policymakers, or the private sector's environmental impact, and the damage caused by each action, as well as the negative effects that these actions can have on affected people in various regions of the world.

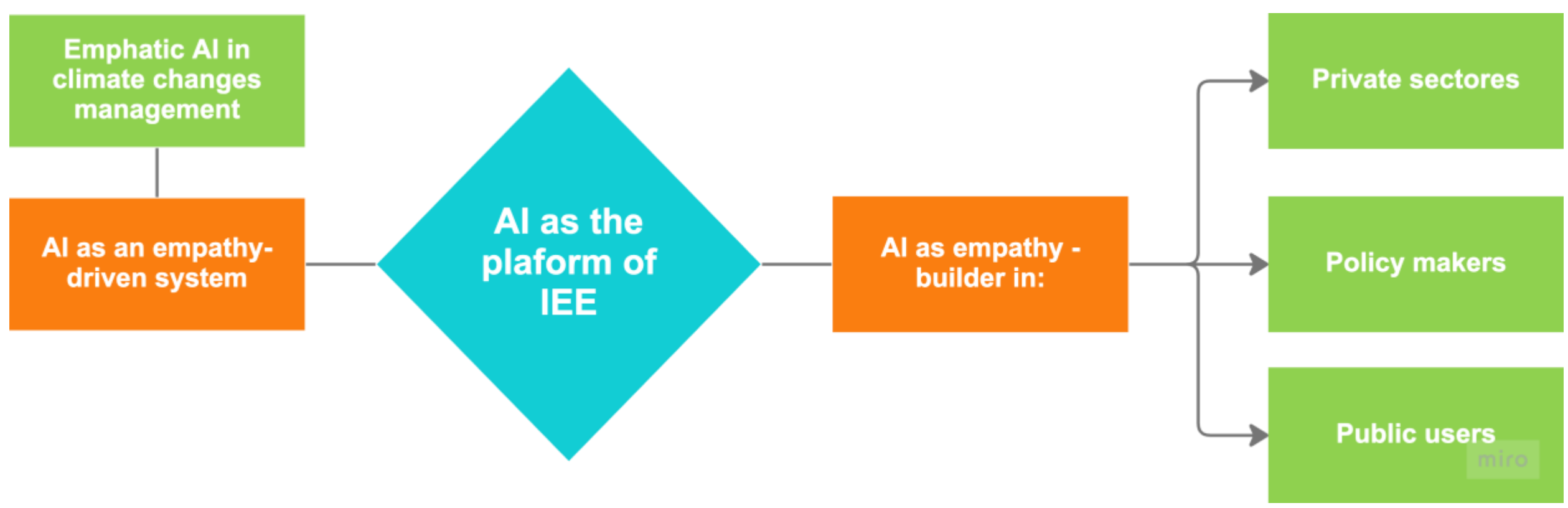

**Figure 4: AI correlation with empathy in IEE**

Such a green movement and motivation would be effective only if it were global enough, and the AI platform might be the sole putative self-controlled global platform that could provide such a possibility to address the global issue of climate injustice. To do that, we need to consider all the following agents as the operational target of green AI systems. That is, we need to design AI systems to create empathy in all the following classes of agents: Global organizations, Traditional states, regional governments, Private international companies, Private logical companies, Individual environmental activists, and Individual citizens. For example, emotional design can be used to create more empathetic AI systems and applications by incorporating design elements that evoke emotional responses in users. Take for example, a system that employs vibrant hues, captivating visuals, and resonating melodies to emulate the repercussions of climate change in a specific locality, thus enlightening users about the ramifications of ecological decline on nearby societies. Furthermore, it is crucial to ascertain that the evolution and execution of AI systems for EEI encompass a wide array of viewpoints and inclusivity, encompassing even those from marginalized communities. This can help ensure that the systems are effective in building empathy among a wide range of users.

The use of AI systems for Environmental Empathy Initiatives (EEI) offers several benefits. One of the most remarkable advantages lies in the capacity to personalize messaging and information to individual users. By customizing information to the specific interests, preferences, and values of each user, AI can heighten engagement and inspire action on environmental issues. Furthermore, AI can unearth patterns and trends in environmental data that humans may not readily discern, yielding invaluable insights into the causes and ramifications of environmental predicaments.

Another pivotal benefit of AI systems is their aptitude to unearth and rectify hidden or elusive environmental injustices. AI can scrutinize data from diverse sources and pinpoint areas where environmental injustices unfold, such as disparate pollution exposure in certain communities. This information can inform policy decisions and help mitigate these injustices, promoting a more just and equitable society.

Additionally, AI can facilitate communication and collaboration among individuals and communities working towards climate justice. By intertwining individuals with akin passions and aspirations, AI has the capability to forge a vibrant and cohesive community, providing unwavering backing for environmental crusades. Moreover, as AI systems evolve into more sophisticated entities, they hold the potential to emulate the enduring repercussions of diverse environmental regulations and initiatives, thereby assisting policymakers and societies in making judicious choices when tackling climate metamorphosis and sundry ecological predicaments. By fostering active participation from communities, AI can guarantee that climate equity resolutions are contrived with invaluable insights from those who shall encounter their ramifications most acutely.

AI, as a global platform with some potential such as being decentralized and having a non-tracking feature like that in Blockchain, could enable and supplement this model with a green surveillance system, which is necessary for such models that are supposed to create and promote green motivation and movement in the global scale [65], [66]. Otherwise, all the attempts might fail due to the problem of free riders. Accordingly, green surveillance would be defined as a designed and developed intelligent surveillance system to track the free riders and data gap regarding environmental issues, and it also requires to be empowered by empathetic design. Green surveillance would be sustainable only it if does not violate other broadly accepted human rights while recognizing cultural and contextual differences and indigenous manners and lifestyles. Such an intelligent surveillance system should not violate other civil rights like privacy and should be contextually sensitive to distinguish between different situations and agents that might entail different evaluations and norms[67], [68], [69]. Moreover, to ensure that AI systems for EEI reach a wide range of users, it is important to address the digital divide, which refers to the unequal access to digital technologies and the internet. This can be done by developing low-cost and accessible AI systems and applications, as well as improving internet access in marginalized communities. Thus, empathic design, with an individual and communal sensitivity feature, seems necessary for an intelligent green surveillance system as well[70].

## 6. Conclusion and future direction

It is broadly agreed that the current top-down approach toward climate justice is not sufficient to reach the putative goals in the next decades. Any complimentary approach needs a green obligation that comes out of the development of environmental motivation in indigenous and human factors in society, the private sector, and governmental policymakers. Such motivators could be external such as UN guidelines and rules that are supported by United Nations Security Council and the military forth which, as we explained, would probably end up in more conflicts in the world, and so would result in a *reductio ad absurdum*. In this paper, focusing on the internal motivation of empathy, we proposed Intelligent Environmental Empathy (IEE) as a new driver for climate peace and justice. The force of green obligation in IEE is not only indigenous and human-centric, but also is empowered by the putative decentralized platform of AI, which enables it to be inclusive enough in dealing with the global problem of climate justice and the free riders problem.

AI systems possess the capacity to wield a momentous influence in tackling environmental challenges by cultivating empathy among citizens, city planners, and administrators. These systems employ AI technologies like machine learning, data analytics, and natural language processing to furnish tailored, instantaneous feedback. In doing so, they enable individuals to apprehend the ramifications of their actions on the environment. Moreover, AI can aid city planners and administrators in formulating more sustainable policies and practices by scrutinizing copious amounts of data, discerning intricate patterns, and making prescient prognostications. While numerous AI systems are currently being developed and implemented to tackle climate and environmental justice, it is of utmost importance to acknowledge that these systems must be meticulously devised and implemented, as they have the ability to perpetuate existing inequalities if executed haphazardly. It is crucial that all stakeholders, including marginalized communities, actively engage in the development process, ensuring that ethical considerations such as bias and privacy are meticulously addressed. With this due regard, AI systems can become potent instruments in forging a more sustainable and equitable world for all.

The future prospects for "intelligent environmental empathy for climate justice" encompass several distinct aspects. Firstly, AI models must attain a higher degree of sophistication to aptly perceive and respond to environmental and social cues. Secondly, let us expand the scope of AI systems to embrace other environmental justice quandaries, like purifying water, enhancing air quality, and managing natural resources. Thirdly, it is of utmost importance to cultivate collaboration amid scientists, policymakers, and local communities in order to augment the caliber and availability of environmental and social data. Fourthly, ethical and privacy concerns pertaining to data collection, analysis, and usage necessitate the formulation of ethical frameworks and guidelines. Lastly, researchers and policymakers ought to engage with local communities to collaboratively devise and implement AI systems that are attuned to their needs and preferences.

An additional factor to ponder is the potential ethical implications of employing AI systems to foster environmental empathy. As AI systems become increasingly advanced, they may acquire the ability to manipulate individuals' emotions and attitudes towards environmental issues. It is imperative to guarantee that the implementation of AI in this context upholds transparency, ethics, and honors individual autonomy. Furthermore, one must also ponder the possibility for AI systems to intensify prevailing power asymmetries and disparities within society. For instance, particular communities or populations may bear a disproportionate burden of environmental issues, and the utilization of AI systems to address these concerns may inadvertently perpetuate these inequalities. It is paramount to contemplate how the development and implementation of AI systems for environmental empathy can be executed in a manner that fosters social justice and equality. Additionally, it is crucial to establish metrics and evaluation frameworks to gauge the efficacy of AI systems and applications in cultivating environmental empathy. This will aid in identifying areas for improvement and ensuring that resources are allocated towards the most efficacious solutions.

Lastly, it is imperative to grasp that AI systems in isolation cannot untangle the convoluted and manifold conundrum of climate justice. While AI can unquestionably exert a pivotal influence in nurturing empathy and expediting well-informed decision-making, it ultimately rests on individuals, communities, and governments to embark on action in alleviating and adjusting to the ramifications of climate change.

**Institutional Review Board Statement:** Not applicable.
**Informed Consent Statement:** Not applicable.
**Data Availability Statement:** Not applicable.

**Conflict of interest:** The authors declare that the research was conducted in the absence of any commercial or financial relationships that could be construed as a potential conflict of interest.
**Acknowledgements:** In accordance with MLA (Modern Language Association) guidelines, we acknowledge the use of AI-powered tools, such as Anthropic's applications, for their assistance and support including in writing, editing, and improving the clarity and organization of the manuscript.

## References

[1] UNEP, "Global Environment Outlook 6," Mar. 2019.
[2] NOAA, "National Oceanic and Atmospheric Administration ," 2015.
[3] P. Hoeppe, "Trends in weather related disasters – Consequences for insurers and society," *Weather Clim Extrem*, vol. 11, pp. 70–79, 2016, doi: 10.1016/j.wace.2015.10.002.

[4] STEPHEN LEAHY, “Most countries aren’t hitting 2030 climate goals, and everyone will pay the price,” *Natl Geogr Mag*, Nov. 2019.

[5] M. B. Bättig and T. Bernauer, “National Institutions and Global Public Goods: Are Democracies More Cooperative in Climate Change Policy?,” *Int Organ*, vol. 63, no. 2, pp. 281–308, 2009, doi: 10.1017/S0020818309090092.

[6] W. Nordhaus, “Climate Clubs: Overcoming Free-Riding in International Climate Policy,” *American Economic Review*, vol. 105, no. 4, pp. 1339–1370, 2015, doi: 10.1257/aer.15000001.

[7] J. Gupta, “Negotiating challenges and climate change,” *Climate Policy*, vol. 12, no. 5, pp. 630–644, 2012, doi: 10.1080/14693062.2012.693392.

[8] D. , P. D. Ghimire, “Interconnection of Climate Change, Agriculture and Climate Justice: Complexities for Feeding the World Under Changing Climate,” *Development*, 2016.

[9] J. Goodman, “From Global Justice to Climate Justice? Justice Ecologism in an Era of Global Warming,” *New Political Science*, vol. 31, no. 4, pp. 499–514, 2009, doi: 10.1080/07393140903322570.

[10] F. Lucivero, “Big Data, Big Waste? A Reflection on the Environmental Sustainability of Big Data Initiatives,” *Sci Eng Ethics*, vol. 26, no. 2, pp. 1009–1030, 2020, doi: 10.1007/s11948-019-00171-7.

[11] M. I. Espinoza and M. Aronczyk, “Big data for climate action or climate action for big data?,” *Big Data Soc*, vol. 8, no. 1, p. 2053951720982032, 2021, doi: 10.1177/2053951720982032.

[12] A. Mah, “Environmental justice in the age of big data: challenging toxic blind spots of voice, speed, and expertise,” *Environ Sociol*, vol. 3, no. 2, pp. 122–133, 2017, doi: 10.1080/23251042.2016.1220849.

[13] “The UN Environment Programme needs new powers,” *Nature*, vol. 591, Mar. 2021.

[14] D. Hume, *A treatise of human nature*. Courier Corporation, 203AD.

[15] John. Locke, *The second treatise of civil government*. Broadview Press, 2015.

[16] T. Skillington, *Climate Change and Intergenerational Justice*. London: Routledge, 2019.

[17] A. V. , and S. E. Burke. Sanson, “Climate change and children: An issue of intergenerational justice,” in *Children and peace*, Springer, 2020.

[18] G. Parietti, “Hobbes on Teleology and Reason,” *Eur J Philos*, vol. 25, no. 4, pp. 1107–1131, 2017, doi: 10.1111/ejop.12244.

[19] N. Gooding and K. Hoekstra, “Hobbes and Aristotle on the Foundation of Political Science,” in *Hobbes’s <I>On the Citizen</I>*, 1st ed., R. Douglass and J. Olsthoorn, Eds., Cambridge University Press, 2019, pp. 31–50. [Online]. Available: https://www.cambridge.org/core/product/identifier/9781108379892%23CN-bp-2/type/book_part

[20] S. Afroogh, Amir Esmalian, Jonan Donaldson, and Ali Mostafavi, “Empathic Design in Engineering Education and Practice: An Approach for Achieving Inclusive and Effective Community Resilience,” *Sustainability*, vol. 13, no. 7, 2021.

[21] I. van de Poel and P. Kroes, “Can Technology Embody Values?,” in *The Moral Status of Technical Artefacts*, P. Kroes and P.-P. Verbeek, Eds., in Philosophy of Engineering and Technology. , Dordrecht: Springer Netherlands, 2014, pp. 103–124. [Online]. Available: https://doi.org/10.1007/978-94-007-7914-3_7

[22] S. Afroogh *et al.*, “Embedded Ethics for Responsible Artificial Intelligence Systems (EE-RAIS) in disaster management: a conceptual model and its deployment,” *AI and Ethics*, Jun. 2023, doi: 10.1007/s43681-023-00309-1.

[23] R. Cohon, “Hume’s Moral Philosophy,” *The Stanford Encyclopedia of Philosophy*, 2018.

[24] M. Mason, “Hume and Humeans on Practical Reason,” *Hume Studies*, vol. 31, no. 2, pp. 347–378, 2005, [Online]. Available: https://muse.jhu.edu/article/383291

[25] J. Mintoff, “Hume and Instrumental Reason,” p. 20, [Online]. Available: files/8639/Mintoff - Hume and Instrumental Reason.pdf

[26] I. Knez, “Is Climate Change a Moral Issue? Effects of Egoism and Altruism on Pro-Environmental Behavior,” *Current Urban Studies*, vol. 4, no. 2, pp. 157–174, 2016, doi: 10.4236/cus.2016.42012.

[27] B. L. Robinson, “Climate Justice: Walter Benjamin and the Anthropocene,” *The Germanic Review: Literature, Culture, Theory*, vol. 96, no. 2, pp. 124–142, 2021, doi: 10.1080/00168890.2021.1897502.

[28] L. et al. Porter, “Climate justice in a climate changed world ,” *Planning Theory & Practice*, vol. 21, no. 2, 2020.

[29] J. Wilkens and A. R. C. Datchoua-Tirvaudey, “Researching climate justice: a decolonial approach to global climate governance,” *Int Aff*, vol. 98, no. 1, pp. 125–143, 2022, doi: 10.1093/ia/iiab209.

[30] D. Falzon and P. Batur, “Lost and Damaged: Environmental Racism, Climate Justice, and Conflict in the Pacific,” in *Handbook of the Sociology of Racial and Ethnic Relations*, P. Batur and J. R. Feagin, Eds., in Handbooks of Sociology and Social Research. , Cham: Springer International Publishing, 2018, pp. 401–412. [Online]. Available: https://doi.org/10.1007/978-3-319-76757-4_22

[31] J. L Rice, J. Long, and A. Levenda, “Against climate apartheid: Confronting the persistent legacies of expendability for climate justice,” *Environ Plan E Nat Space*, vol. 5, no. 2, pp. 625–645, 2022, doi: 10.1177/2514848621999286.

[32] H. Hassani, X. Huang, and E. Silva, “Big Data and Climate Change,” *Big Data and Cognitive Computing*, vol. 3, no. 1, p. 12, 2019, doi: 10.3390/bdcc3010012.

[33] G. A. Veltri and D. Atanasova, “Climate change on Twitter: Content, media ecology and information sharing behaviour,” *Public Understanding of Science*, vol. 26, no. 6, pp. 721–737, 2017, doi: 10.1177/0963662515613702.

[34] H.-D. Guo, L. Zhang, and L.-W. Zhu, “Earth observation big data for climate change research,” *Advances in Climate Change Research*, vol. 6, no. 2, pp. 108–117, 2015, doi: 10.1016/j.accre.2015.09.007.

[35] J. H. Faghmous and V. Kumar, “A Big Data Guide to Understanding Climate Change: The Case for Theory-Guided Data Science,” *Big Data*, vol. 2, no. 3, pp. 155–163, 2014, doi: 10.1089/big.2014.0026.

[36] Alexander Saier, “Greater Climate Ambition Urged as Initial NDC Synthesis Report Is Published,” *UN Climate Change Conference*, Feb. 2021.

[37] N. Roughley, “The Empathy in Moral Obligation,” in *Forms of Fellow Feeling: Empathy, Sympathy, Concern and Moral Agency*, 2018.

[38] J. Decety and J. M. Cowell, “Empathy, Justice, and Moral Behavior,” *AJOB Neurosci*, vol. 6, no. 3, pp. 3–14, 2015, doi: 10.1080/21507740.2015.1047055.

[39] J. Berenguer, “The Effect of Empathy in Proenvironmental Attitudes and Behaviors,” *Environ Behav*, vol. 39, no. 2, pp. 269–283, 2007, doi: 10.1177/0013916506292937.

[40] J. Berenguer, "The Effect of Empathy in Environmental Moral Reasoning," *Environ Behav*, vol. 42, no. 1, pp. 110–134, 2010, doi: 10.1177/0013916508325892.

[41] S. EVANS, "Which countries are historically responsible for climate change?," *Carbon Brief*, 2021.

[42] H. Bauder and R. Mueller, "Westphalian Vs. Indigenous Sovereignty: Challenging Colonial Territorial Governance," *Geopolitics*, pp. 1–18, 2021, doi: 10.1080/14650045.2021.1920577.

[43] J. Havercroft, "Was Westphalia 'all that'? Hobbes, Bellarmine, and the norm of non-intervention," *Global Constitutionalism* , vol. 1, no. 1, 2012.

[44] T. Hobbes and Marshall Missner, *Thomas Hobbes: Leviathan* . Routledge, 2016.

[45] D. Moellendorf, "Climate change and global justice," *WIREs Climate Change*, vol. 3, no. 2, pp. 131–143, 2012, doi: 10.1002/wcc.158.

[46] J. D. Petersen-Perlman, J. C. Veilleux, and A. T. Wolf, "International water conflict and cooperation: challenges and opportunities," *Water Int*, vol. 42, no. 2, pp. 105–120, 2017, doi: 10.1080/02508060.2017.1276041.

[47] B. H. Weston, "Climate Change and Intergenerational Justice: Foundational Reflections," *Vermont Journal of Environmental Law*, vol. 9, p. 375, 2007, [Online]. Available: https://heinonline.org/HOL/Page?handle=hein.journals/vermenl9&id=381&div=&collection=

[48] C. Motupalli, "International Justice, Environmental Law, and Restorative Justice," *Wash J Environ Law Policy*, vol. 8, p. 333, 2018, [Online]. Available: https://heinonline.org/HOL/Page?handle=hein.journals/wshjoop8&id=339&div=&collection=

[49] A. Carrapatoso, "In Search of Alternative Governance Models—The Contribution of Interregional Climate Cooperation to the Global Climate Change Regime," *Earth System Governance Conference*, 2012.

[50] M. J. , et al. Brueggemann, "Reflexive Practices for the Future of Design Education: An Exercise in Ethno-Empathy," *The Design Journal* , vol. 20, 2017.

[51] J. Decety and K. J. Yoder, "Empathy and motivation for justice: Cognitive empathy and concern, but not emotional empathy, predict sensitivity to injustice for others," *Soc Neurosci*, vol. 11, no. 1, pp. 1–14, 2016, doi: 10.1080/17470919.2015.1029593.

[52] C. Huntingford, E. S. Jeffers, M. B. Bonsall, H. M. Christensen, T. Lees, and H. Yang, "Machine learning and artificial intelligence to aid climate change research and preparedness," *Environmental Research Letters*, vol. 14, no. 12, p. 124007, 2019, doi: 10.1088/1748-9326/ab4e55.

[53] P. S. et al. Yeung, "Investigating Future Urbanization's Impact on Local Climate under Different Climate Change Scenarios in MEGA-urban Regions: A Case Study of the Pearl River Delta, China," *Atmosphere (Basel)*, vol. 11, no. 7, 2020.

[54] L. A. Mansfield, P. J. Nowack, M. Kasoar, R. G. Everitt, W. J. Collins, and A. Voulgarakis, "Predicting global patterns of long-term climate change from short-term simulations using machine learning," *NPJ Clim Atmos Sci*, vol. 3, no. 1, pp. 1–9, 2020, doi: 10.1038/s41612-020-00148-5.

[55] P. Voosen, "Europe builds 'digital twin' of Earth to hone climate forecasts," *Science (1979)*, vol. 370, no. 6512, pp. 16–17, 2020, doi: 10.1126/science.370.6512.16.

[56] O. Folorunsho and A. Adeyemo, “Application of Data Mining Techniques in Weather Prediction and Climate Change Studies,” *International Journal of Information Engineering and Electronic Business*, vol. 4, 2012, doi: 10.5815/ijieeb.2012.01.07.

[57] Y. Takao, “Making climate change policy work at the local level: capacity-building for decentralized policy making in Japan,” *Pac Aff*, vol. 85, no. 4, 2012.

[58] K. H. L. and K. C. Lo, “Climate experimentation and the limits of top-down control: local variation of climate pilots in China,” *Journal of Environmental Planning and Management*, vol. 63, no. 1, 2020.

[59] S. Andresen, “International Climate Negotiations: Top-down, Bottom-up or a Combination of Both?,” *The International Spectator*, vol. 50, no. 1, pp. 15–30, 2015, doi: 10.1080/03932729.2014.997992.

[60] A. Goldthau, “Rethinking the governance of energy infrastructure: Scale, decentralization and polycentrism,” *Energy Res Soc Sci*, vol. 1, pp. 134–140, 2014, doi: 10.1016/j.erss.2014.02.009.

[61] F. Tormos-Aponte and G. A. García-López, “Polycentric struggles: The experience of the global climate justice movement,” *Environmental Policy and Governance*, vol. 28, no. 4, pp. 284–294, 2018, doi: 10.1002/eet.1815.

[62] et al. Grimaud A, “Climate change mitigation options and directed technical change: A decentralized equilibrium analysis,” *Resource and Energy Economics*, vol. 33, no. 4, 2011.

[63] Y. et al. Cheng, “How do technological innovation and fiscal decentralization affect the environment? A story of the fourth industrial revolution and sustainable growth,” *Technological Forecasting and Social Change* , vol. 162, 2021.

[64] J. Kruger, W. E. Oates, and W. A. Pizer, “Decentralization in the EU Emissions Trading Scheme and Lessons for Global Policy,” *Rev Environ Econ Policy*, 2007, doi: 10.1093/reep/rem009.

[65] Y.-P. Lin, J. R. Petway, W.-Y. Lien, and J. Settele, “Blockchain with Artificial Intelligence to Efficiently Manage Water Use under Climate Change,” *Environments*, vol. 5, no. 3, p. 34, 2018, doi: 10.3390/environments5030034.

[66] R. Sivarethinamohan, P. Jovin, and S. Sujatha, “Unlocking the Potential of (AI-Powered) Blockchain Technology in Environment Sustainability and Social Good,” in *Applied Edge AI*, Auerbach Publications, 2022.

[67] J. Jiao, S. Afroogh, Y. Xu, and C. Phillips, “Navigating LLM Ethics: Advancements, Challenges, and Future Directions,” *arXiv*. [Online]. Available: http://arxiv.org/abs/2406.18841

[68] S. Afroogh *et al.*, “Tracing app technology: an ethical review in the COVID-19 era and directions for post-COVID-19,” vol. 24, no. 3, p. 30, doi: 10.1007/s10676-022-09659-6.

[69] S. Afroogh, A. Akbari, E. Malone, M. Kargar, and H. Alambeigi, “Trust in AI: Progress, Challenges, and Future Directions,” *arXiv*. [Online]. Available: http://arxiv.org/abs/2403.14680

[70] S. Afroogh and A. Esmalian, “Empathic Design for Community Resilience”, [Online]. Available: https://philpapers.org/rec/AFREDF